\documentclass[journal=ancham,manuscript=article]{achemso}

\usepackage{chemformula} % Formula subscripts using \ch{}
\usepackage[T1]{fontenc} % Use modern font encodings
\usepackage{amsmath}
\usepackage{amssymb}
\usepackage{upgreek}
\usepackage[symbol]{footmisc}
\usepackage{color}
\usepackage{float}
\usepackage{wasysym}
\usepackage[version=3]{mhchem}
\usepackage{booktabs,dcolumn,caption}
\usepackage{nicematrix}
\usepackage{multirow}
\usepackage{booktabs}
\usepackage{makecell}
\usepackage{enumitem}
\usepackage{textgreek}
\usepackage{caption}
\usepackage{subcaption}
\usepackage{graphicx}

\usepackage{xr}
\usepackage{cleveref}
\newcolumntype{d}[1]{D{.}{.}{#1}} 
\renewcommand{\thefootnote}{\roman{footnote}}

\mciteErrorOnUnknownfalse

\DeclareRobustCommand{\rchi}{{\mathpalette\irchi\relax}}
\newcommand{\irchi}[2]{\raisebox{\depth}{$#1\chi$}}

\SectionNumbersOn

\title{Piezo-to-Piezo (P2P) Conversion: Simultaneous $\beta$-Phase Crystallization and Poling of Ultrathin, Transparent and Freestanding Homopolymer PVDF Films via MHz-Order Nanoelectromechanical Vibration} 

\author{Robert Komljenovic}
\affiliation{Micro/Nanophysics Research Laboratory, School of Engineering, RMIT University, Melbourne, VIC 3001, Australia}
\author{Peter C.~ Sherrell}
\affiliation{Applied Chemistry and Environmental Science, School of Science, RMIT University, Melbourne, VIC 3001, Australia}
\author{Amgad R.~Rezk}
\affiliation{Micro/Nanophysics Research Laboratory, School of Engineering, RMIT University, Melbourne, VIC 3001, Australia}
\email{amgad.rezk@rmit.edu.au}
\author{Leslie Y.~Yeo}
\affiliation{Micro/Nanophysics Research Laboratory, School of Engineering, RMIT University, Melbourne, VIC 3001, Australia}


\begin{document}

\renewcommand{\thefootnote}{\fnsymbol{footnote}}

\pagebreak
\begin{abstract}
\sloppy

An unconventional yet facile low-energy method for uniquely synthesizing neat poly(vinylidene fluoride) (PVDF) films for energy harvesting applications through piezo-to-piezo (P2P) conversion is reported. In this novel concept, the nanoelectromechanical energy from a piezoelectric substrate is directly coupled into another polarizable material (i.e., PVDF) during its crystallization to produce a micron-thick film that not only exhibits strong piezoelectricity, but is also freestanding and optically transparent---properties ideal for its use for energy harvesting, but which are difficult to achieve through conventional synthesis routes. In particular, we show that the unprecedented acceleration ($\mathcal{O}$($10^{8}$ m s$^{-2}$)) associated with the nanoelectromechanical vibration in the form of surface reflected bulk waves (SRBWs) facilitates preferentially-oriented nucleation of the ferroelectric PVDF $\beta$-phase, while simultaneously aligning its dipoles to pole the material through the SRBW's intense native evanescent electric field ($\mathcal{O}$($10^{8}$ V m$^{-1}$)). The resultant neat (additive-free) homopolymer film synthesized through this low voltage method requiring only  $\mathcal{O}$(10 V)---orders-of-magnitude lower than the energy-intensive conventional poling methods utilising high kV electric potentials---is shown to possess a 76\% higher macroscale piezoelectric charge coefficient ($d_{33}$), together with a similar improvement in its power generation output, when compared to the gold-standard commercially-poled PVDF films of similar thicknesses.

\fussy
\end{abstract}

\pagebreak

\section{Introduction}

Bulk inorganic ferroelectric ceramics, such as lead zirconium titanate (PbZrTiO$_{3}$; PZT) or barium titanate (BaTiO$_{3}$; BTO), have been extensively studied as potential materials for energy harvesting applications \cite{vallem2021energy,fan2016flexible,shi2018implantable,park2014highly} given their superior piezoelectric properties \cite{li1991extrinsic,buscaglia2004ferroelectric,bowen2014piezoelectric}. Their use in practice is, however, limited by their rigidity and brittleness, and the need for solid-state sintering during their fabrication \cite{shi1999high,vijayakanth2022recent}. Organic piezoelectric polymers, such as poly(vinylidene fluoride) (PVDF), on the other hand, are thin, lightweight, flexible, durable and biocompatible\cite{lu2020flexible,de2022sustainable}. Consequently, they have been widely explored as  piezoelectric alternatives to bulk ceramic materials for flexible energy harvesting \cite{soin2015exclusive,shepelin2021interfacial, song2021polyvinylidene, huang2021enhanced,su2022high} and wearable sensing \cite{islam2023boosting,kim20173d,wang2012progress}, amongst other applications \cite{sun2023over,wang20222,chen2017table}. Nevertheless, despite PVDF having superior piezoelectric properties among the range of organic polymers known to date \cite{lopes2018direct,martins2014electroactive,lu2020flexible}, the ability to produce freestanding, neat (additive-free), transparent and ultrathin homopolymer PVDF films with high piezoelectricity remains challenging. \cite{huang2022ultrarobust,li2022piezoelectric} 

In general, the ability to produce highly piezoelectric PVDF relies on (1) preferential crystal orientation of its ferroelectric $\beta$-phase, and, (2) subsequent $\beta$-phase dipole alignment (i.e., poling). To achieve the first, a PVDF film, which is usually composed predominantly of the non-ferroelectric $\alpha$-phase in its natural state, is traditionally heated (to approximately 80 $^{\circ}$C) and concurrently stretched, either uni- or bi-axially, often at high strain (e.g. 400\% stretch ratios), \cite{ting2018characteristic} to induce a higher $\beta$-phase fraction. \cite{yin2016effects,wang2018electroactive, tang2020effects,shehata2022stretchable,salimi2004conformational} Given the propensity for the films to crack and tear, however, such mechanical methods for $\alpha$- to $\beta$-phase conversion tend to be limited to relatively thick films (> 10 $\upmu$m) \cite{ting2018characteristic,tao2022fused}. Alternatively, greater $\beta$-phase fractions can be obtained by conjugating PVDF with other polymers (for example, poly(trifluoroethylene); TrFE), or through the addition of nanofillers (such as graphene oxide, metal-organic frameworks (MOFs), two-dimensional titanium carbides/nitrides (MXenes), piezoceramic nanoparticles and other materials).\cite{wang2022mxenes,cao2021piezoelectric,roy2020three} Compared to homopolymers, however, copolymers are significantly more expensive (the cost of PVDF-TrFE, for example, is more than tenfold that of PVDF),\cite{gryshkov2021pvdf} while the addition of nanofillers leads to non-uniform distribution in the $\beta$-phase crystallization, in which the $\beta$-phase is typically co-localised in the filler regions \cite{islam2023boosting}.

For the latter, i.e., $\beta$-phase poling, two broad strategies have been employed to pole the film: the application of a large electric field that surpasses the coercive field of the neat (additive-free) homopolymer PVDF film, or by exploiting the self-poling effect of the nanofillers that are added to the film  \cite{karan2016approach,nardekar2022mos2}. On the one hand, the use of high electric fields---ranging from $10^{7}$--$10^{8}$ V m$^{-1}$ (through methods such as corona, electrospinning, or electrode/contact poling)\cite{tao2022fused,chung2022enhanced,huang2022enhanced}---has long been shown to successfully result in polarized neat PVDF. Such approaches, however, not only necessitate the consumption of large amounts of energy, costly infrastructure and high kV sources \cite{bao2023flexible,fan2022effect}, but can also often lead to dielectric breakdown, particularly for thin films. \cite{fan2020improving,jain2015dielectric} In contrast, the incorporation of fillers provides for a low-energy alternative, but compromises the transparency and flexibility of the film. \cite{zhang2023recent,chatterjee2023nanofiller,nunes2018poly} 

From this perspective, a facile, one-step, low-energy method for synthesizing thin freestanding neat (additive-free) homopolymer PVDF films possessing a large fraction of molecularly-oriented and poled ferroelectric $\beta$-phase that leads to high levels in its piezoelectricity remains unrealised. In this work, we report such a method for the first time; a piezo-to-piezo (P2P) conversion mechanism in which a piezoelectric substrate (in this case, lithium niobate; LiNbO$_3$) is harnessed to facilitate {\em in situ} simultaneous crystallization and poling of a polarizable piezoelectric (albeit, one that is weak, to begin with) homopolymer (in this case, PVDF) to produce a highly piezoelectric ultrathin ($\upmu$m-order) freestanding film. More specifically, we show that the unique nanoelectromechanical coupling \cite{rezk2021high} associated with hybrid surface and bulk acoustic waves (i.e., surface reflected bulk waves, or SRBWs) \cite{rezk2016hybrid} generated on the LiNbO$_3$ substrate gives rise to an extraordinary surface acceleration ($\mathcal{O}$($10^{8}$ m s$^{-2}$)) and electric field ($\mathcal{O}$($10^{8}$ V m$^{-1}$)) \cite{rezk2020free} capable of simultaneously inducing---in a single step---both the mechanical stimulation required to facilitate formation of the $\beta$-phase together with the electric field necessary to induce alignment of its dipoles, concurrently during its crystallization. The result is the production of an ultrathin film with high $\beta$-phase fraction ($F_{\beta}$ = 74\%), surpassing that for both the solution-cast control ($F_{\beta}$ = 42\%) and a gold-standard commercially-poled sample ($F_{\beta}$ = 58\%), and yielding a 76$\%$ increase in the macroscale piezoelectric charge coefficient ($d_{\rm 33}$) compared with commercially-poled film of comparable thickness. As such, the P2P SRBW platform removes the need for the cumbersome, lengthy and energy-intensive processes associated with conventional post-synthesis phase conversion and electrical poling, constituting an efficient, novel and green alternative for the synthesis of high performance freestanding PVDF films.

\pagebreak

\section{Results and Discussion}

\begin{figure}
\centering
\includegraphics[width=0.8\textwidth]{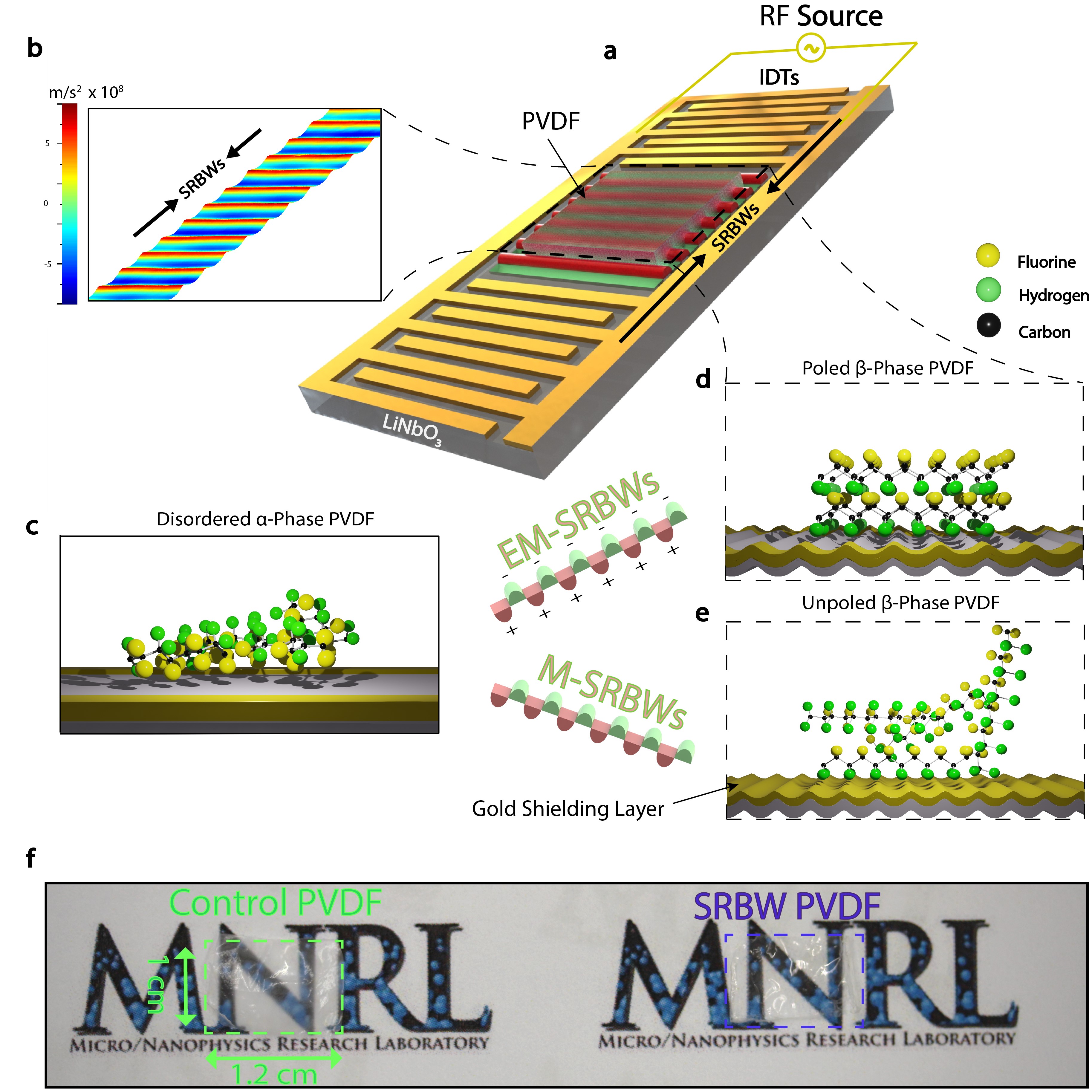}
\caption{(a) Schematic of the SRBW device employed for simultaneous synthesis and poling of ultrathin PVDF films. The electromechanical vibration in the form of SRBWs are generated by applying a RF signal in the form of an oscillating electrical signal at the resonant frequency of the device (10 MHz) to a pair of interdigitated transducers (IDTs) photolithographically patterned on the lithium niobate (LiNbO$_3$) substrate. The SRBWs that are launched from each IDT can be seen to propagate along and through the substrate in opposing directions. As a consequence, they superimpose to form a standing wave beneath the region where the PVDF precursor solution is deposited and where the PVDF film forms, as indicated by (b) the laser Doppler vibrometer scan of the substrate vibration acceleration in this region. Compared to (c) the control experiment in which PVDF is conventionally synthesized in the absence of the SRBW forcing, yielding a film that mainly comprised a disordered non-polar $\alpha$-phase, the electromechanical coupling associated with the SRBW (EM-SRBW) resulted in (d) a film with an oriented  polar $\beta$-phase. If a gold film is patterned on the LiNbO$_3$ substrate to short out the electric field, the synthesis of the PVDF film in the presence of a predominantly mechanical stress field (M-SRBW) only led to a (e) primarily disordered $\beta$-phase film. Doing so allows us to isolate the crucial role of the evanescent SRBW electric field during the synthesis and poling. (f) The SRBW-synthesized PVDF film (EM-SRBW) can be seen to be more optically transparent than the control PVDF film of similar thickness.}
\label{figure1}
\end{figure}

The experimental setup and protocol for the synthesis and characterization of the PVDF films are illustrated in Fig.~\ref{figure1}(a) and described in the Methods section, respectively. In particular, we compare PVDF films synthesized by drop casting precursor solutions of PVDF powder dissolved in acetone and N,N-dimethylformamide (DMF) onto the LiNbO$_3$ substrate (Fig.~\ref{figure1}(a)), both in the absence of the SRBW excitation but under heating to $100\ ^\circ$C for 30 min as the control (Fig.~\ref{figure1}(c)), and in the presence of the SRBW excitation (Fig.~\ref{figure1}(b)) and hence the full electromechanical coupling (EM-SRBW; Fig.~\ref{figure1}(d)) at a power of 15 dBm for 20 min and subsequently 30 dBm for a further 20 min. To elucidate the role of the electromechanical coupling, and more specifically the mechanical and electric fields separately, we repeated the synthesis in the presence of the SRBW excitation, but without its native electric field such that the crystallization of the precursor solutions into the PVDF film occurred while being subjected solely to the mechanical vibration (M-SRBW; Fig.~\ref{figure1}(e)). This was achieved by depositing a thin gold shielding layer atop the substrate on which the precursor solutions were deposited in order to screen out the evanescent electric field. In all of the cases, the precursor solutions were observed to crystallize into ultrathin films during the solvent evaporation process, whose $\upmu$m-order thickness was controlled primarily by adjusting the volume of the precursor solution deposited. Considerable differences were however observed with regards to the molecular orientation and dipole alignment of the phases, as well as the physical properties (in particular, the piezoelectricity as well as the optical transmittance (Fig.~\ref{figure1}(f))), between the control and SRBW-synthesized films.

\pagebreak
\subsection{Crystal Structure and Surface Morphology: SRBW Mechanical Stress Facilitates $\beta$-Phase Formation} 
\label{sec:alphabeta}

\begin{figure}
%\centering
\includegraphics[width=0.85\columnwidth]{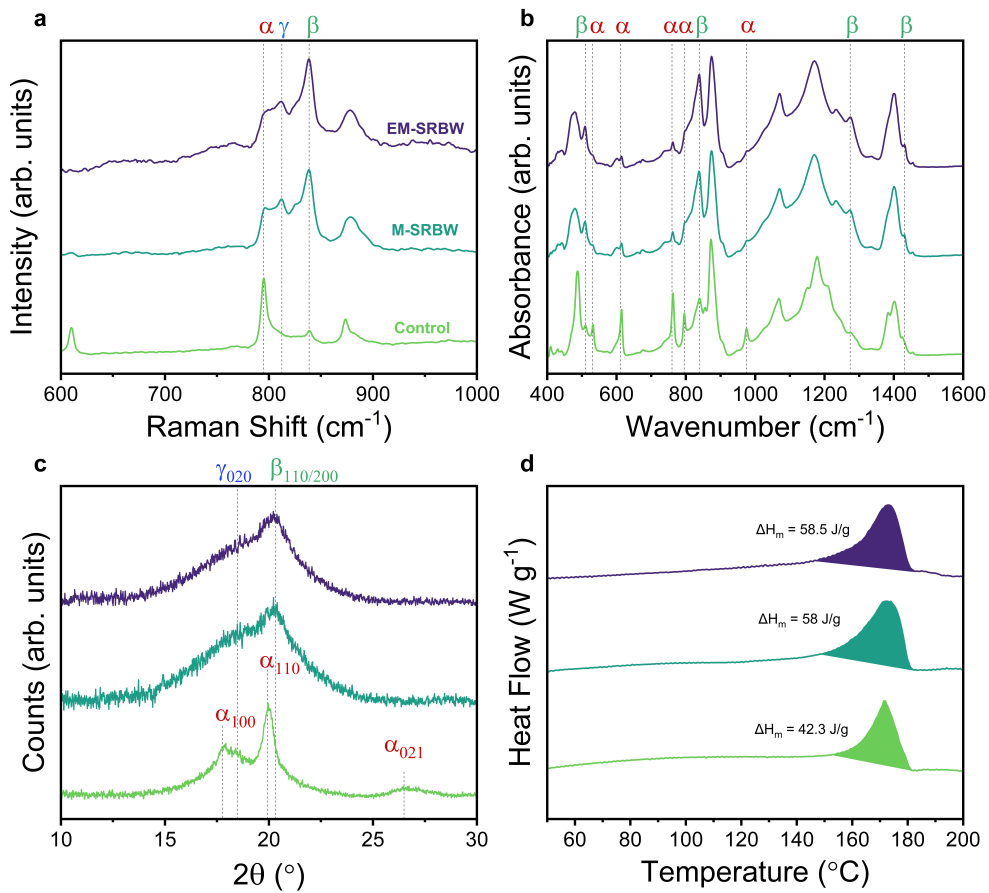}
\caption{(a) Raman, (b) Fourier-transform infrared (FTIR), and (c) powder x-ray diffraction (XRD) spectra, and, (d) differential scanning calorimetry (DSC) heat trace for the control and SRBW-synthesized PVDF films, both in the absence (M-SRBW) and presence (EM-SRBW) of the SRBW evanescent electric field. In (a), (b) and (c), the relevant peaks for the $\alpha$-, $\beta$- and $\gamma$-phases are shown.}
\label{figure2}
\end{figure}

Figure \ref{figure2}(a), shows the Raman spectra for the PVDF films synthesized via solution casting (control) and via SRBW synthesis, both in the absence (pure mechanical stress; M-SRBW) and presence (full electromechanical coupling; EM-SRBW) of its evanescent electric field. It can be seen that all of the synthesized films showed the characteristic peaks associated with the $\alpha$- and $\beta$-phases for PVDF at 794 and 839 cm$^{-1}$, respectively, that have been suggested to result from the rocking of their corresponding CH$_2$ bonds \cite{kobayashi1975molecular,shepelin2019new}. Nevertheless, we note the prominence of the $\beta$-phase peak and suppression of the $\alpha$-phase peak, together with the appearance of the peak associated with the electroactive $\gamma$-phase at 812 cm$^{-1}$ for both SRBW-synthesized samples. More specifically, the relative intensity ratio of the $\beta$- to $\alpha$-phases ($I_\beta/I_\alpha$) can be seen to increase from 0.31 for the control PVDF film to 1.86 and 1.82 for the EM-SRBW and M-SRBW films, respectively, alluding that exposure of the PVDF film to the SRBW during its synthesis leads to suppression of the non-ferroelectric $\alpha$-phase and the preferential crystallization of the ferroelectric $\beta$-phase. We note that these values surpass even that for the state-of-the-art 3D-printed PVDF-MoS$_2$ ($I_\beta/I_\alpha$ $\approx$ 0.95)\cite{islam2023boosting} and are comparable to those reported for Ti$_3$C$_2$T$_x$/PVDF-TrFE composites ($I_\beta/I_\alpha$ = 2.5) \cite{shepelin2021interfacial}.

Moreover, Raman mapping (Fig.~S\ref{Sfigure2}) over a relatively large area ($100\ \upmu$m $\times 100\ \upmu$m) further demonstrates the homogeneity in the distribution of the $\beta$-phase across the films subjected to the SRBW exposure, in contrast to the non-uniform distributions observed for PVDF nanocomposites that incorporate fillers to induce $\beta$-phase crystallization, in which the $\beta$-phase is typically co-localised in the filler regions. \cite{islam2023boosting} We conducted a brief parametric study to obtain the optimum SRBW input power that results in the largest $\beta$-phase molecular fraction (Fig.~S\ref{Sfigure3}), demonstrating that this is an extremely efficient process to promote $\beta$-phase formation even with the minimal powers used.

The increase in  $\beta$-phase content in the SRBW excited films is further verified through an analysis of the Fourier Transform Infrared (FTIR) spectra in Fig.~\ref{figure2}(b) for each of the synthesized PVDF samples, noting that peaks at 530, 612, 763, 795 and 973 cm$^{-1}$ are associated with the $\alpha$-phase and peaks at 510, 841, 1276 and 1429 cm$^{-1}$ are associated with the $\beta$-phase.\cite{cai2017critical,mirejamethods,makarevich1965infrared} The Beer-Lambert law (Eq.\ref{Eq1}) then allows an estimation of the proportion of the $\beta$-phase content in the samples:\cite{su2021muscle,salimi2003analysis}
\begin{equation} \label{Eq1}
F_{\beta} = {\frac{A_{\beta}}{(1.26A_{\alpha} + A_{\beta})}},
\end{equation}
wherein $A_\alpha$ and $A_\beta$ indicate the intensities of the FTIR peaks at 763 and 841 cm$^{-1}$, respectively; the factor of 1.26 for the $\alpha$-phase compensates for the difference in the absorption coefficients, which are 6.1 × 10\,$^{4}$ and 7.7 × 10\,$^{4}$ cm$^{2}$ mol$^{-1}$ for the $\alpha$- and $\beta$-phases, respectively \cite{gregorio1994effect}. Both the EM-SRBW and M-SRBW films were found to possess $F_{\beta}$ values of 74.4\% and 72.7\% respectively, which is equivalent to an approximate 76\% increase in $\beta$-phase fraction over the control ($F_{\beta}$ = 42.1\%), and a 28\% increase over the gold-standard commercially-poled sample of comparable thickness ($F_{\beta}$ = 58\%; Fig.~S\ref{Sfigure4}).

The effectiveness of the mechanical stresses generated along the piezoelectric substrate by the SRBW on the PVDF sample in selectively inducing the $\beta$-phase, rather than the typical $\alpha$-phase, are further corroborated by the powder x-ray diffraction (XRD) spectra in the results above. From Fig.~\ref{figure2}(c), it can be seen that the control film, whose diffraction peaks are located at $17.7^\circ$, $19.9^\circ$ and $26.6^\circ$, corresponding to the (100), (110) and (021) reflections of the monoclinic $\alpha$-phase crystal, respectively, appear to be suppressed in both the SRBW-synthesized films.\cite{lei2013spectroscopic,wu2015facile}. In their place, we observed peaks at $18.5^\circ$ for the (020) plane corresponding to the monoclinic $\gamma$-phase, in addition to more prominent peaks at 20.4$^\circ$, corresponding to the (110) and (200) planes commonly attributed to the orthorhombic $\beta$-phase,\cite{yin2019characterization,cai2017effect,martins2014electroactive} which have typically been assigned within the range 20.26$^\circ$ to 20.6$^\circ$. \cite{gregorio2006determination,shepelin2019new,szewczyk2020enhanced,taleb2021fabrication}. 

Both the M-SRBW and EM-SRBW PVDF films furthermore display broader and larger areas ($\Delta H_m = 58$--$58.5$ J g$^{-1}$) below the melting peak ($T_m = 170^\circ$) on their respective differential scanning calorimetry (DSC) curves when compared to the more narrow area of the control film ($\Delta H_m = 42.3$ J g$^{-1}$), as seen in Fig.~\ref{figure2}d. This is a consequence of the $\beta$-phase having a lower-temperature state,\cite{szewczyk2020enhanced, shepelin2019new} driven by stronger polar interactions within its all-trans (TTT) planar zigzag conformations \cite{he2022poly}. Besides highlighting the prevalence of the $\beta$-phase in the SRBW-synthesized films, the DSC results also show an  appreciable increase in the overall crystallinity ($\rchi_{c}$) of these films, which can be calculated from 
\begin{equation} \label{Eq2}
	\rchi_{c} = {\frac{{{\Delta}H_m}}{{\Delta}H^0_m}\times100\%},
\end{equation}
where $\Delta H_m$ and  $\Delta H^0_m$ denote the enthalpy of melting for the PVDF film that is synthesized and that for purely crystalline PVDF ($\Delta H^0_m = 104.5$ J g$^{-1}$),\cite{pusty2022defect,zheng2022structure} respectively. The calculated $\rchi_{c}$ value for the M-SRBW and EM-SRBW synthesized films were 59.9\% and 60.5\%, respectively, which are considerably higher than that for the control film ($\rchi_{c}$ = 44\%).

\begin{figure}
%\centering
\includegraphics[width=\columnwidth]{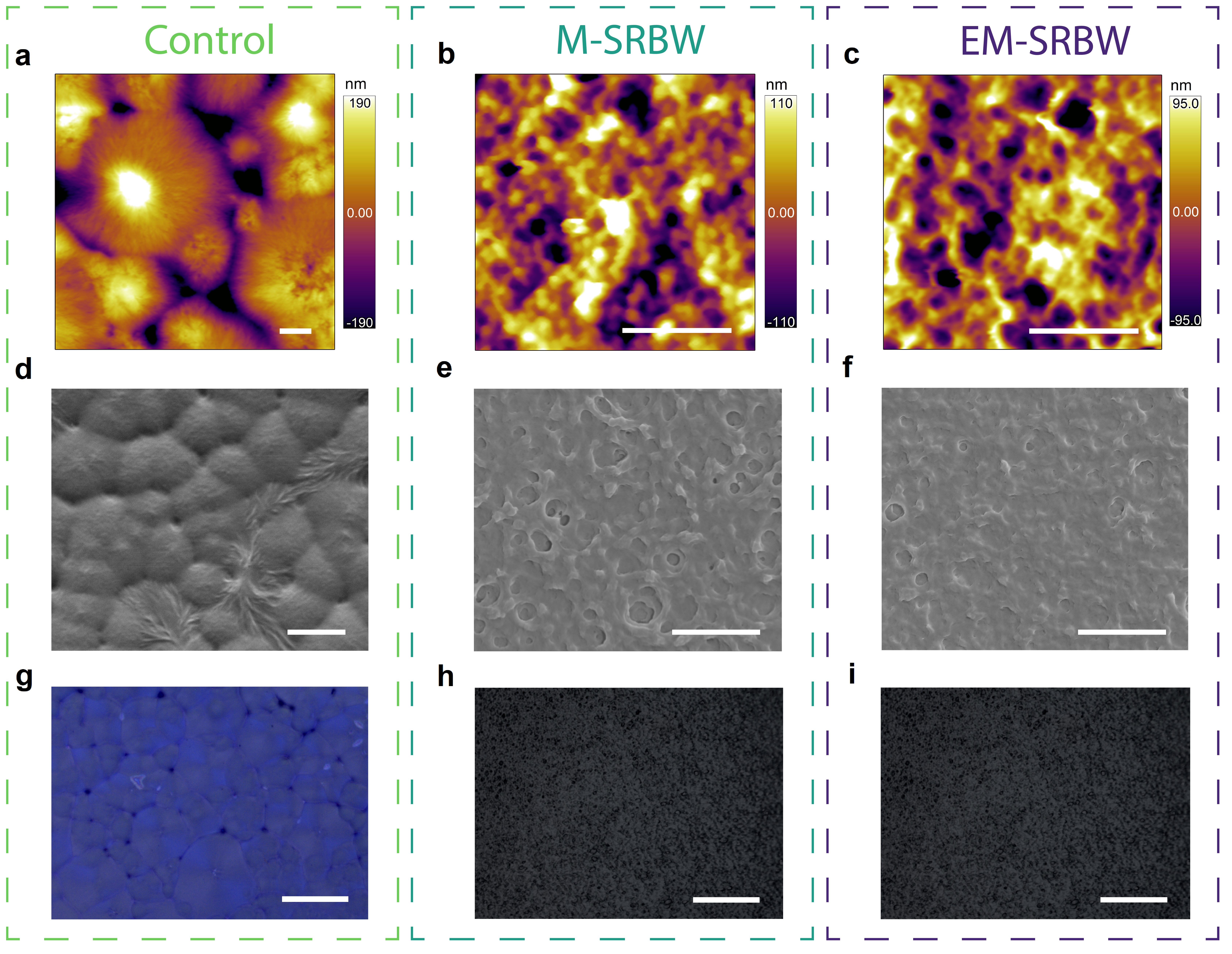}
\caption{(a--c) Atomic force microscopy (AFM; scale bars: $2\ \upmu$m), (d--f) scanning electron microscopy (SEM; scale bars: $2\ \upmu$m), and, (g--i) polarized optical microscopy (POM; scale bars: $20\ \upmu$m) images showing the surface morphology of the PVDF film synthesized via (a,d,g) solution casting (control; left box), and, by the SRBW forcing, both (b,e,h) in the absence (M-SRBW; centre box) and (c,f,i) in  the presence (EM-SRBW; right box) of the SRBW evanescent electric field.}
\label{figure3}
\end{figure}

The characterization of the samples, as summarised in Table 1, collectively suggest that the SRBW-synthesized films are predominantly composed of the $\beta$-phase, compared to the $\alpha$-phase dominant control films. This can further be seen from the morphology of the films, captured by the microscopy images in Fig.~\ref{figure3}. In particular, we observe the SRBW-synthesized films (Fig.~\ref{figure3}(b,c,e,f,h,i)) to be devoid of the typical well-defined and regular ringed spherulitic structure that appear as Maltese extinction crosses under polarized optical microscopy, characteristic of the $\alpha$-phase, \cite{sencadas2009alpha,sencadas2012relaxation, li2014studies,miyashita2022two} prominent in the control films (Fig.~\ref{figure3}(a,d,g)). 

\begin{table}[ht]
\setlength\tabcolsep{22.5pt}
\centering
\caption{Degree of crystallinity ($\rchi_{c}$) from DSC measurements, ${\beta}$-phase ($F_{\beta}$) and ${\alpha}$-phase ($F_{\alpha}$) compositions from FTIR spectroscopy, and the relative $\beta$- to $\alpha$-phase fractions $I_{\beta}/I_{\alpha}$ obtained from Raman spectroscopy for each of the PVDF films synthesized.}
\begin{tabular}[t]{lcccc}
\hline
& $\rchi_{c}$ (\%)
& $F_{\beta}$ (\%)
&$F_{\alpha}$ (\%)
& $I_{\beta}/I_{\alpha}$\\
\hline
Control & 43.8 & 42.1 & 52.9 & 0.31 \\
M-SRBW & 59.9 & 74.4 & 25.6 & 1.82 \\
EM-SRBW& 60.5 & 72.7 & 27.3 & 1.86 \\
\hline
\end{tabular}
\end{table}

The ability of the SRBW forcing to suppress $\alpha$-phase growth and to allow preferential nucleation of the $\beta$-phase during crystallization of the polymer can be understood from the unique nanomechanical interactions arising at the solid--liquid interface of the piezoelectric LiNbO$_3$ substrate, along which the SRBW propagates \cite{rezk2021high}. Despite being only several nanometers in amplitude, the high frequency of the SRBW (10 MHz) yields surface accelerations that are exceptionally large ($\mathcal{O}$($10^{8}$ m s$^{-2}$)). It is thus likely that the local dynamical stress variations---on the order of several MPa \cite{massahud2023acoustomicrofluidic}---that the SRBW imparts on the PVDF film as it crystallizes act to introduce large numbers of nucleation sites throughout the film, in a manner akin to sonocation-induced nucleation in the sonocrystallization of polymers (such as poly-3-hexylthiophene) \cite{lee2016physicochemical,xi2018sonocrystallization}, although this has never been demonstrated for PVDF. These local nucleation sites, in turn, act to disrupt the stable growth of $\alpha$-phase spherulites, in a manner similar to the way the electrostatic interactions arising from the presence of nanofillers hinder $\alpha$-phase growth and promote proliferation of the $\beta$-phase (we note though that in the typically non-uniform distribution of the nanofillers, however, tend to result in $\beta$-phases that are non-uniformly distributed in these cases). \cite{yuan2022poling,roy2020three,martins2011nucleation} The SRBW platform thereby effectively allows the formation of the irregular, textual variations comprising considerably smaller protrusions associated with $\beta$-phase PVDF seen in Fig.~\ref{figure3}(b,c,e,f); this transformation and hindered crystalline structure being further evidenced by the low birefringence observed in the polarized light microscopy images in Fig.~\ref{figure3}(h,i), which is indicative of $\beta$-phase PVDF \cite{song2020effect}.

The morphological differences between the films were also observed to influence their macroscopic properties. A side-by-side visual comparison of the control and EM-SRBW films in Fig.~\ref{figure1}(f), together with their total optical transmittance across the visible spectrum (380--700 nm; Fig.~S\ref{Sfigure1}) shows that the EM-SRBW films possess higher transparency (approximately 90\% transmittance) than the control films, and a similar transmittance to that for commercially-acquired films of similar thicknesses (6 $\upmu$m). This can be attributed to the higher $\beta$-phase composition, particularly given that the spherulitic $\alpha$-phase superstructure has been noted to contribute to the opacity of the film.
\cite{osaka2013optical}

\subsection{Local Polarization Effects: SRBW Evanescent Electric Field Drives Simultaneous Poling}
\label{sec:pole}

\begin{figure}
\centering
\includegraphics[width=\columnwidth]{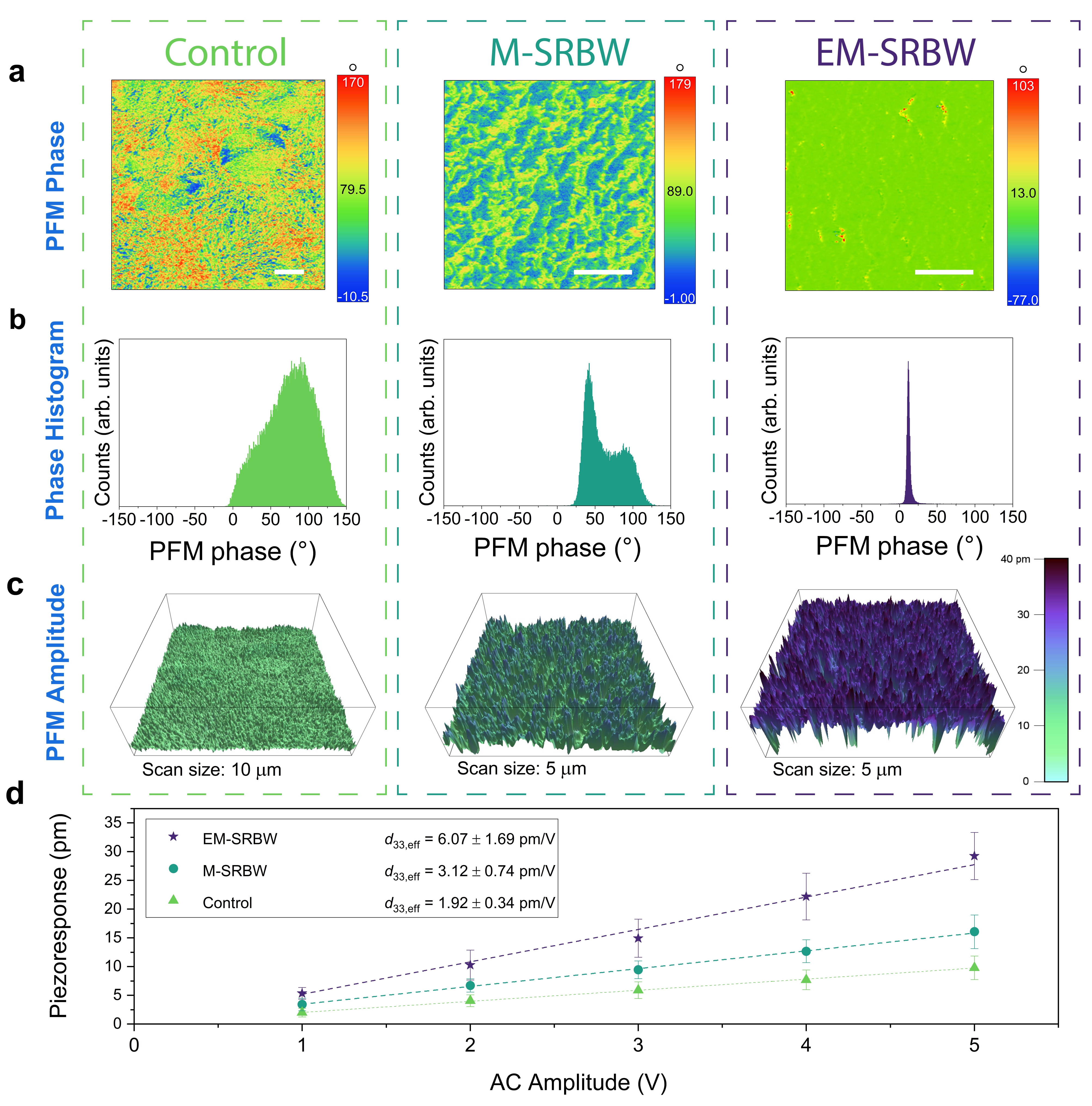}
\caption{Piezoelectric force microscopy (PFM) (a) phase, (b) phase distribution, and, (c) piezo-response amplitude (at 5 V$_{\rm AC}$) for the control (left box) and SRBW-synthesized films, both in the absence (M-SRBW; centre box) and in the presence (EM-SRBW; right box) of the SRBW evanescent electric field. (d) Piezoelectric force response and piezoelectric coefficient ($d_{33,{\rm eff}}$) for each of the films. The scale bars in (a) denote lengths of 2 $\upmu$m.}
\label{figure4}
\end{figure}

To demonstrate the effect of the evanescent electric field associated with the SRBW nanoelectromechanical coupling on the poling of the $\beta$-phase, we now quantify the local piezoelectric response in the synthesized films using piezoelectric force microscopy (PFM). In particular, the central role of the SRBW evanescent electric field can immediately be seen from the PFM phase profile in Fig.~\ref{figure4}(a,b), which characterises the local polarization direction, i.e., the dipole orientation of the PVDF molecules within the specified area of interest \cite{soin2016high,kuang2021piezoelectric}. More specifically, we observe a broad phase distribution that characterises predominantly disordered polarization in the control film, which does not appreciably change with the SRBW synthesis in the absence of its evanescent electric field (M-SRBW), wherein the width of the corresponding phase histogram exhibits a similarly broad range. It is only when this electric field is present that there is a significant narrowing of the phase distribution to signify highly-oriented local polarization direction \cite{pariy2019piezoelectric,ferri2019local}, thereby alluding to the critical role of the SRBW evanescent electric field in simultaneously poling the material concurrently during its synthesis (for comparison, a similarly narrow phase distribution was obtained for commercially-acquired PVDF film that has been post-synthetically poled (Fig.~S\ref{Sfigure5}(a,b)).

The PFM amplitude for the various PVDF films that were synthesized is shown in Fig.~\ref{figure4}(c), and is directly related to the piezoelectric response in Fig.~\ref{figure4}(d), whose slope (with respect to the AC voltage applied between the PFM tip and the local unit cell surface along the PVDF film) quantifies the piezoelectric coefficient ($d_{33,{\rm eff}}$).  It can thus be seen that there is an increase in $d_{33,{\rm eff}}$ of approximately 65\% when the PVDF is synthesized solely under the SRBW mechanical stress (M-SRBW), from 1.92 pm V$^{-1}$ for the control to 3.12 pm V$^{-1}$, which can be attributed to the greater proportion of $\beta$-phase in the sample discussed above, despite it being predominantly disordered, as shown in Fig.~\ref{figure4}(a,b). The $d_{33,{\rm eff}}$ value, nevertheless, further increases to 6.07 pm V$^{-1}$ for the EM-SRBW film when the full electromechanical coupling associated with the SRBW is present---a significant 216\% increase over the control film, which we conclude from Fig.~\ref{figure4}(a,b) to arise due to the poling effect of the SRBW evanescent electric field in aligning the dipoles to give rise to a predominantly ordered $\beta$-phase simultaneously during its crystallization within the film. This value is superior, not only to that obtained with the commercially-poled 
film of comparable thickness (5.26  pm V$^{-1}$; Fig.~S\ref{Sfigure5}(c)) but also the $d_{33,{\rm eff}}$ values which have been reported for PVDF films poled via electrospinning involving high kV potentials (5.56  pm V$^{-1}$) \cite{szewczyk2020enhanced}, and that for
composite Ti$_3$C$_2$T$_x$/PVDF-TrFE employing considerably more costly MXene nanofillers (5.11  pm V$^{-1}$) \cite{shepelin2019new}. 

The role of the evanescent electric field component of the SRBW in locally aligning the dipoles of the $\beta$-phase induced by its mechanical component is not unexpected in light of its intensity\cite{rezk2020free}---on the order 10$^8$ V m$^{-1}$---compared with typical electric field intensities that have been reported for poling PVDF ($3 \times 10^{7}$ -- $65 \times 10^{7}$ V m$^{-1}$)\cite{tao2022fused,zhu2021electrostriction,huang2021enhanced}. Such molecular orientation and dipole rearrangement under the SRBW is not without precedent: the SRBW, for example, has previously been shown to induce similar dipole alignment during the crystallization of MOFs to result in highly-oriented structures \cite{ahmed2019acoustomicrofluidic}. The P2P EM-SRBW platform is unique (compared to post-synthesis poling methods utilising high electric fields) given its ability not only to pole the PVDF film simultaneously during crystallization, but to also require substantially lower applied voltages ($\approx$ 1--10 V) compared to the kV voltages typical of electrical poling techniques \cite{bao2023flexible,fan2022effect}. The ability of the SRBW to pole the PVDF film at these substantially lower applied voltages is possible because of the electric field confinement over the nanometer lengthscales associated with the SRBW displacement amplitude \cite{rezk2020free}.

\pagebreak

\subsection{Macropiezoelectricity Measurements}

\begin{figure}
\centering
\includegraphics[width=1.05\columnwidth]{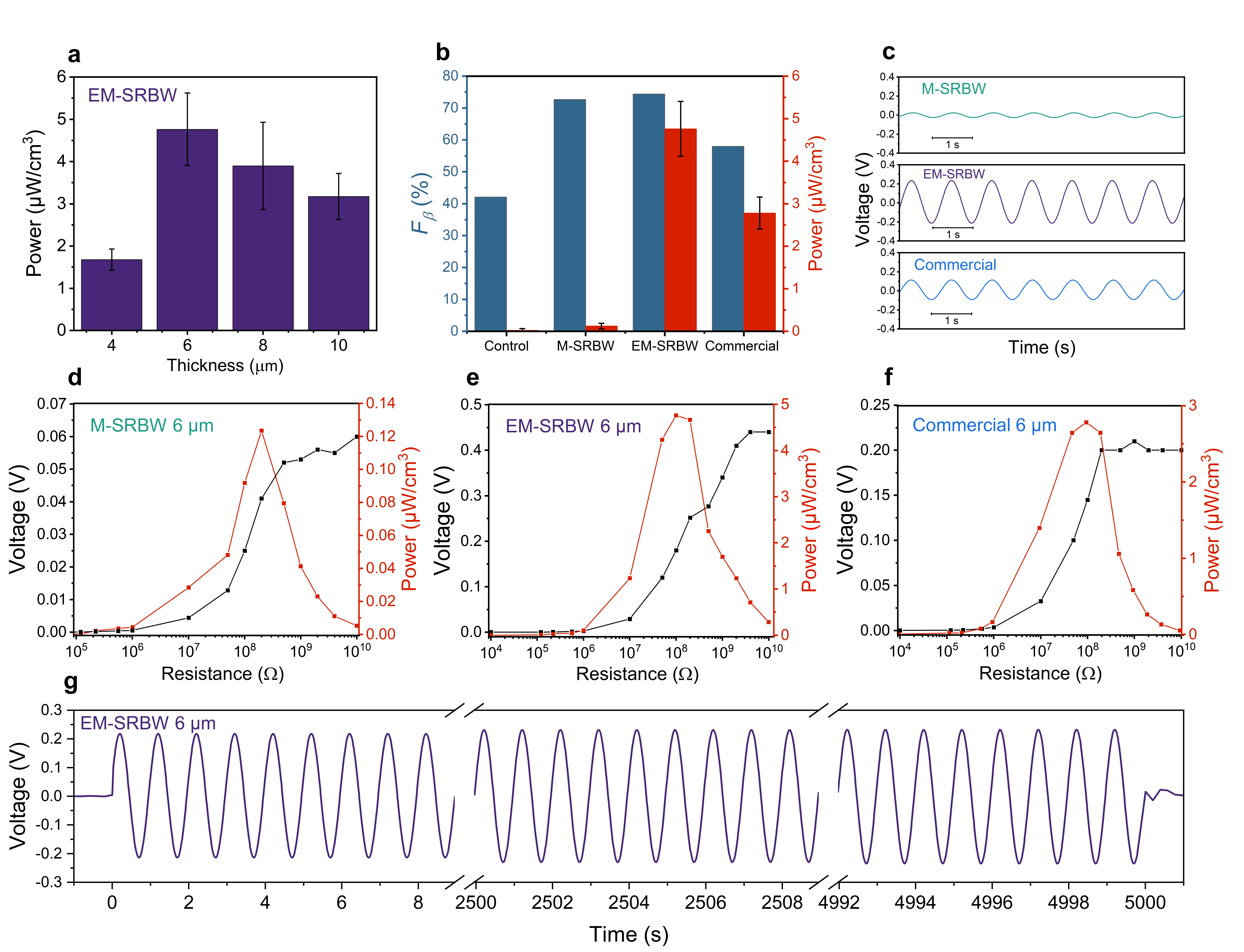}
\caption {Macroscale evaluation of control, M-SRBW, EM-SRBW and commercially-poled PENG devices subjected to 1 Hz sinusoidal cyclic \emph{in-contact} compression forces (5--15 N). Power density (\textit{P$_{\rm D}$}) output (a) for EM-SRBW
PENG devices with varying PVDF film thicknesses, and, (b) as compared across different PENG devices with the same film thickness ($6\ \upmu$m); their corresponding open-circuit voltages being shown in (c). (d--f) Voltage and power density output as a function of  resistive load for the different PENG devices. (g) Voltage output for the EM-SRBW PENG device ($6\ \upmu$m) over multiple compression cycles at the open-circuit voltage.}
\label{figure5}
\end{figure}

To evaluate the device-scale macroscopic piezoelectric properties of the control and SRBW-synthesized films in comparison to those for commercially-poled PVDF films of similar thicknesses, we constructed piezoelectric nanogenerator (PENG) devices for each of the films by coating them with Cr/Au electrodes and encasing them with insulating polyimide tape, as shown in (Fig.~S\ref{Sfigure6}), full details for which are given in the Methods section. The power output for the EM-SRBW PENG devices for different thickness films when subjected to a 1 Hz sinusoidal cyclic \emph{in-contact} compressive force (5--15 N with 10 N preloading to minimise artefacts associated with contact electrification (i.e., contact separation, charge induction and lateral-sliding triboelectric modes)\cite{shepelin2021interfacial} that can often inflate the piezoelectric output) \cite{vsutka2020measuring,chen2022method} is shown in Fig.~\ref{figure5}(a), indicating optimum power generation with PVDF films of 6$\ \pm 1\ \upmu$m thickness. Importantly, to demonstrate that the obtained voltage outputs solely arose as a consequence of the film's piezoelectricity, we conducted Fast Fourier Transform (FFT) analyses of the time domain waveforms associated with the input signals. In particular we note that a spectral bandwidth of 8 Hz, as determined through frequency domain FFT analysis of signals obtained from 0.11 mm contact separation measurements related to the film's triboelectric effects (see Fig. S\ref{Sfigure7}(a)), reduces to 1 Hz for measurements taken while in contact, which are exclusively linked to piezoelectric effects (see Fig. S\ref{Sfigure7}(b)) in the EM-SRBW device. Moreover, the bandwidth for in-contact measurements being lower than $2f_{0}$ indicates the absence of substantial triboelectric electrostatic interference mechanisms affecting the overall voltage output. \cite{leon2023decoupling}

The optimal power output at a film thickness of 6 $\upmu$m (Fig.~\ref{figure5}(a)) is likely due to reduced dipole alignment in thinner ($< 6\ \upmu$m) PVDF films as a consequence of the free surface instabilities arising as a result of the acoustic radiation pressure imposed by the underlying SRBW forcing on the air--liquid interface when the initial film height of the liquid that forms following the spreading of the precursor solution as it is dispensed onto the LiNbO$_3$ substrate, prior to its crystallization, is below $100\ \upmu$m. Weaker dipole alignment is also the reason for the reduced power output in thicker  ($> 6\ \upmu$m) PVDF films, but this instead being due to the decrease in penetration of the SRBW evanescent electric field into the initial liquid film, when its thickness exceeds the SRBW wavelength (approximately $400\ \upmu$m).

Subsequently, we compared the voltage and power output of devices with a thickness of $6\ \upmu$m, produced using the SRBW platform, with commercially poled films of similar thickness  (Fig.~\ref{figure5}(b)--\ref{figure5}(f)). A twofold increase in the peak-to-peak open-circuit voltage is observed for the EM-SRBW PENG device (0.4 V) in comparison to the 0.2 V obtained for the commercially-poled device (Fig.~\ref{figure5}(e,f)). The power density output, on the other hand, can be seen to increase with the resistive load, reaching an optimum at an impedance of 100 M\textOmega\ (200 M\textOmega\ for the M-SRBW PENG device). In any case, it can be seen from Fig.~\ref{figure5}(b) that the EM-SRBW PENG device, having a greater $\beta$-phase composition and having been poled by the SRBW's evanescent electric field, delivers significantly higher voltage and power output compared to both the control and M-SRBW PENG devices. The latter having a greater $\beta$- to $\alpha$-phase fraction than the control but lacking alignment in its dipoles and hence resulting in overall cancellation of the dipole vectors throughout the film \cite{shepelin2019new}, due to the absence of the SRBW electric field; the correlation between the $\beta$-phase fraction for each film and the power output clearly being seen. Notably, the EM-SRBW PENG device produced a power output (4.76 $\upmu$W cm$^{-3}$) and absolute polarization charge coefficient (12.7 pC N$^{-1}$; (Fig.~S\ref{Sfigure8})) that is more than 70\% greater than that for the commercially-poled PENG device (2.78 $\upmu$W cm$^{-3}$ and 7.2 pC N$^{-1}$, respectively). Moreover, we show the EM-SRBW PENG device to be stable, at least up to 5,000 continuous cycles of operation, in which no appreciable fluctuations in the open-circuit voltage was observed (Fig.~\ref{figure5}(g)). 

Finally, to exemplify the use of the EM-SRBW synthesized films, we briefly demonstrate their superior properties for electronic charging. For these experiments, our samples exhibited impressive capacitor charging speed to 2 V when one capacitor (1, 2 and 10 $\upmu$F) was used, achieved through a simple finger press onto the device with the aid of a bridge rectifier (AC-to-DC converting circuit), as shown in Fig.~S\ref{Sfigure9}(a).  We note a faster charging speed of the device (2 V in approximately 120 s) to that for the commercially-poled PENG device (1.6 V in approximately 200 s; Fig.~S\ref{Sfigure9}(b)), comparable to that observed with state of the art polymer PENGs, \cite{li2023high,song20233d} thereby underscoring the potential of the P2P EM-SRBW PENG devices to harness human motion for powering remote and portable electronics.

\section{Conclusion}

A novel one-step route that exploits the nanoelectromechanical interactions between one piezoelectric substrate into another without an intermediary material (P2P) for synthesizing  PVDF films that possess high levels of $\beta$-phase and are simultaneously poled during its crystallization to yield a material with superior piezoelectricity is presented. In particular, we harness the extraordinary acceleration---on the order of 10 million $g$'s---together with the intense native electric field  of nanoelectromechanical vibrations in the form of SRBWs to simultaneously enable oriented crystallization and align the dipoles during crystallization of the material into an ultrathin ($\upmu$m-thick) freestanding film. The large mechanical stresses associated with the $\mathcal{O}$($10^{8}$ m s$^{-2}$) SRBW substrate acceleration are shown to hinder growth of the spherulites associated with the non-ferroelectric PVDF $\alpha$-phase and to facilitate large numbers of nucleation sites for the formation of the ferroelectric $\beta$-phase, whose dipoles are concurrently aligned by the $\mathcal{O}$($10^{8}$ V m$^{-1}$) SRBW evanescent electric field. 

These synergistic effects are observed to yield freestanding neat micron-thick PVDF films with high $\beta$-phase fractions and which are simultaneously poled---requisite characteristics for a material with strong piezoelectric properties---without the need for additives (e.g., nanofillers) or energy-intensive processes. By fabricating piezoelectric nanogenerator (PENG) devices from these films, we show the films synthesized with this method possess superior properties in terms of optical transparency and piezoelectric charge coefficient ($d_{33}$), the latter translating into $>70$\% improvement in power generation as an energy harvesting device compared to gold-standard commercially-poled films of similar thicknesses. Altogether, this low-voltage and low-cost green approach circumvents conventional energy-intensive processes, in addition to eliminating the need for costly nanofiller materials used in more recent approaches, for the preparation of highly piezoelectric PVDF films. 

\pagebreak

\newpage 
\section*{Methods}
\label{method}

\subsection{PVDF precursor solution} 
To prepare the PVDF precursor solution, we dissolved 3 wt.\% PVDF powder (M$_{m}$ = 238,000, M$_{w}$ = 573,000 g mol$^{-1}$, polydispersity index = 2.4; Solvay Interox Pty.~Ltd., Banksmeadow, NSW, Australia) in 60 vol.\% acetone  ($\ge 99.8\%$, CAS 67-64-1, ACS Reagent; Sigma-Aldrich Pty.~Ltd., Castle Hill, NSW, Australia)  and 40 vol.\% N,N-dimethylformamide (DMF; CAS 68-12-2, ACS Reagent; Sigma-Aldrich Pty.~Ltd., Castle Hill, NSW, Australia) via ultrasonication for 60 min at room temperature. A commercially-poled PVDF film (6 $\upmu$m thickness) was also acquired from PolyK Technologies LLC (State College, PA, USA) for the performance comparison.

\subsection{SRBW platform}
SRBW devices (Fig.~\ref{figure1}(b)) were fabricated by photolithographically patterning a pair of interdigitated transducer (IDT) electrodes, each with 68 finger pairs of aperture 13.94 mm, onto a single crystal piezoelectric LiNbO$_{3}$ substrate (Roditi International Corp.~Ltd., London, UK), The IDTs comprised 10 $\upmu$m titanium (Ti) and  200 $\upmu$m gold (Au) metal layers, whose finger width and gap of $\lambda/4$ specify the SRBW wavelength $\lambda = $ 378 $\upmu$m and hence frequency $f = \lambda/c_s = 10$ MHz, wherein $c_s = $ 3780 ms$^{-1}$ is the acoustic wave phase speed in LiNbO$_{3}$. To generate the SRBW, an RF signal at the resonant frequency (10 MHz) from a signal generator (SML01; Rhode \& Schwarz GmbH \& Co. KG, Munich, Germany) and amplifier  (ZHL-5W-1+; MiniCircuits, Brooklyn, NY, USA) is supplied to the IDTs. The substrate displacement, velocity and acceleration of the SRBW was measured with the aid of a laser Doppler vibrometer (LDV; UHF-120; Polytec Inc., Irvine, CA, USA).

\subsection{PVDF film synthesis} 
Control films were synthesized by drop casting 0.04--0.1 ml (depending on the requisite thickness) of the aforementioned PVDF precursor solution onto the LiNbO$_{3}$ substrate, but without SRBW excitation, and heating to 100 $^{\circ}$C for 30 min. For the SRBW synthesis, the same volume was pipetted onto the LiNbO$_{3}$ substrate and the device actuated initially at a power of 15 dBm for 20 min and subsequently at 30 dBm for a further 20 min, to subsequently yield a film which could then be peeled off the substrate. 

\subsection{Film characterization}
\textbf{Powder x-ray diffraction (XRD)} (D8 General Area Detector Diffraction System (GADDS); Bruker Pty. Ltd., Preston VIC, Australia) was carried out to determine the phases of the control and SRBW-synthesized PVDF films. The analysis was performed at 40 mA and 40 kV Cu-K$\alpha$ radiation ($\lambda$ = 1.54 \AA ) over a $20^{\circ}$--$30^{\circ}$ $2\theta$ range with a step size of $0.01^{\circ}$ and scan rate of $2.6^{\circ}$ min$^{-1}$.\\
\textbf{UV-Vis spectra} (Apollo; CRAIC Technologies, San Dimas, CA, USA) were obtained in the visible wavelength range (between 380 and 700 nm) at a step size of 0.8 nm.\\
\textbf{Raman spectroscopy} (LabRAM HR Evolution; Horiba Scientific SAS, Palaiseau, France) was conducted at 532 nm excitation (600--1000 cm$^{-1}$ acquisition range) with a $100\times$ objective and 1800 g mm$^{-1}$ grating. All spectra were calibrated against   a silicon wafer to a wavelength of 520 cm$^{-1}$. \\
\textbf{Fourier-Transform infrared (FTIR)} (Spectrum One, PerkinElmer Inc., Waltham, MA, USA) transmittance spectra were captured by utilising a compression technique at room temperature across a wide range of 500–4000 cm$^{-1}$ over 64 scans and at high resolution (4 cm$^{-1}$). \\
\textbf{Differential scanning calorimetry (DSC)} (Pyris 1; PerkinElmer, Pontyclun, UK) was employed to determine the crystallinity of the  PVDF films. 5 mg samples were placed in a metal pan and heated to 200 $^{\circ}$C at a ramp rate of $10 ^{\circ}$ min$^{-1}$. \\
\textbf{Scanning electron microscopy (SEM)} (Nova NanoSEM 450, FEI, Hillsboro, OR, USA) imaging was conducted under a 30 kV electron beam with a spot size of 3.5. \\
\textbf{Polarized optical micrographs (POM)} were acquired using an optical microscope (2500; Leica Microsystems GmbH, Wetzlar, Germany)  with a high definition digital camera (DFC290; Leica Microsystems GmbH, Wetzlar, Germany).

\subsection{Local polarization measurements}

Piezoresponse force microscopy (PFM) (Asylum Research MFP-3D Infinity; Oxford Instruments, Santa Barbara, CA, USA) using Dual AC Resonance Tracking Piezo Force Microscopy (DART-PFM) contact mode with a contact resonance frequency shift of approximately 260 kHz was employed for the local piezoelectric measurements. Data acquisition and analysis was conducted using the open source software IGOR Pro (v6.32A; WaveMetrics Inc., Lake Oswego, OR, USA). A conductive Pt/Ir coated atomic force microscopy (AFM) tip (SCM-PIT-V2, resonance frequency 75 kHz, spring constant 3 Nm$^{-1}$, radius $\approx 25$ nm; Bruker Pty. Ltd., Preston VIC, Australia) was used for the DART-PFM measurements. The PVDF sample was adhered using carbon tape to a conductive Au/Cr substrate, which was grounded to the PFM stage prior to the measurement, in which scans were conducted at a frequency of 1 Hz over atleast a $5\ \upmu$m $\times\ 5\ \upmu$m area at 256 pixels per line. An AC driving amplitude is swept from 1 to 5 V whilst the tip was in-contact with the PVDF sample to calculate the quantitative vertical piezoelectric coefficient of the PVDF films, referred to as the effective piezoelectric constant ($d_{33,{\rm eff}}$). Unlike DART \emph{hysteresis} PFM, which frequently experiences the influence of electrostatic interactions during polarization switching, particularly in the ON-field state, \cite{kim2016ferroelectric,liu2020variation,qiao2019electrostatic,qiao2020electrostatic} our approach involved utilising a DC voltage bias to nullify the surface potential of the PVDF and thus reduce the impact of electrostatic effects during DART \emph{scanning} PFM \cite{kim2017electrostatic}. To accurately mitigate and correct for DART scanning electrostatic effects, we conducted local Kelvin probe force microscopy scans to find the local surface potential and thus apply an opposing DC voltage prior to the PFM measurements to offset the surface potential \cite{kim2017electrostatic}. In addition, we also conducted pre-measurements to calculate the $d_{33,{\rm eff}}$ of periodically-polarized LiNbO$_3$ (PPLN; Asylum Research; Oxford Instruments, Santa Barbara, CA, USA) to further ensure reliability and accuracy in the measurement.

\subsection{Macroscale piezoelectricity quantification}

PENG devices ((Fig.~S\ref{Sfigure6}(b,c)) were first fabricated by applying electrodes onto each of the control, SRBW-synthesized and commercially-poled PVDF films via electron beam deposition (PRO Line PVD 75; The Kurt J. Lesker Company, Jefferson Hills, PA, USA). This process involved depositing a 10 nm Cr layer and a 100 nm Au layer onto the films. A shadow mask with a 0.56 cm$^{2}$ active area was first used to define the electrode placement on both sides of the material. Copper foil tape was then attached as a point of contact to establish a solid connection between the tape and the Cr/Au coating, following which wires were soldered onto the tape on both sides of the films to ensure proper contact. Finally, insulating polyimide tape (Kapton\textsuperscript{\textregistered}; DuPont Company, Wilmington, DE, USA) was applied to both surfaces to fully enclose the films.

Macroscale polarization measurements of each PENG device then involved subjecting them to \emph{in-contact} cyclic compressive force and measuring the output voltages. Briefly, a sinusoidal force at 1 Hz frequency with a preload of 10 N and a minimum and maximum load of approximately 5 and 15 N ($\Delta F = 10$ N), respectively, was applied to the active area of the PENG using the dynamic testing instrument (Electropuls E3000; Instron, Norwood, MA, USA) shown in Fig.~S\ref{Sfigure6}(a). The electrical outputs, determined by connecting a known variable resistor (1 k\textOmega--10 G\textOmega) in parallel with the PENG device, were measured using a source meter unit (B2912A; Keysight Technologies, Mulgrave, VIC, Australia) at 0.1 s intervals. From these measurements, the power output density of the device can then be calculated from
\begin{equation} \label{Eq3}
P_{\rm D} = \frac{V_{\rm pp}^2}{A_e R t },
\end{equation}
wherein $V_{\rm pp}$ is the peak-to-peak voltage, $R$ the load resistance, $t$ the film thickness, and $A_e$ the effective surface area of the cylindrical impactor that is in contact with the surface of the device (0.126 cm$^2$), respectively.  To delineate between the individual contributions arising from triboelectric and piezoelectric effects, we also analysed the data in the frequency domain using Fast Fourier Transforms (FFT) \cite{levine1965inverse}. To show the ability of the PENG devices to utilise their piezoelectric effect to convert finger tapping vibrations into electrical energy, a bridge rectifier circuit was used to convert the generated AC voltage into a DC signal, enabling the charging of a capacitor, which was then recorded using an oscilloscope (RTO 1044, Rhode \& Schwarz, North Ryde, NSW, Australia). Discharging subsequently occurs as the stored energy is gradually released from the storage element. A piezoelectric meter (PKD3-2000; PolyK Technologies LLC, State College, PA, USA) with a 2 N static load was separately used to determine the absolute piezoelectric charge coefficient ($d_{\rm 33}$). All experiments were carried out at room temperature.  

\pagebreak

\section*{Conflicts of interest}

There are no conflicts to declare.

\begin{acknowledgement}

The authors are grateful for access to, and for the technical assistance (particularly that of Dr Chenglong Xu) associated with, the use of the equipment and facilities in the RMIT School of Science, the RMIT MicroNano Research Facility (MNRF) and the RMIT Microscopy \& Microanalysis Research Facility (RMMF). 

\end{acknowledgement}

\pagebreak

\bibliography{refrob2}

\end{document}

\renewcommand{\thefootnote}{\fnsymbol{footnote}}

\pagebreak

\begin{figure}[H]
\centering
\hspace*{-1.5cm}
\includegraphics[width=1.2\textwidth]{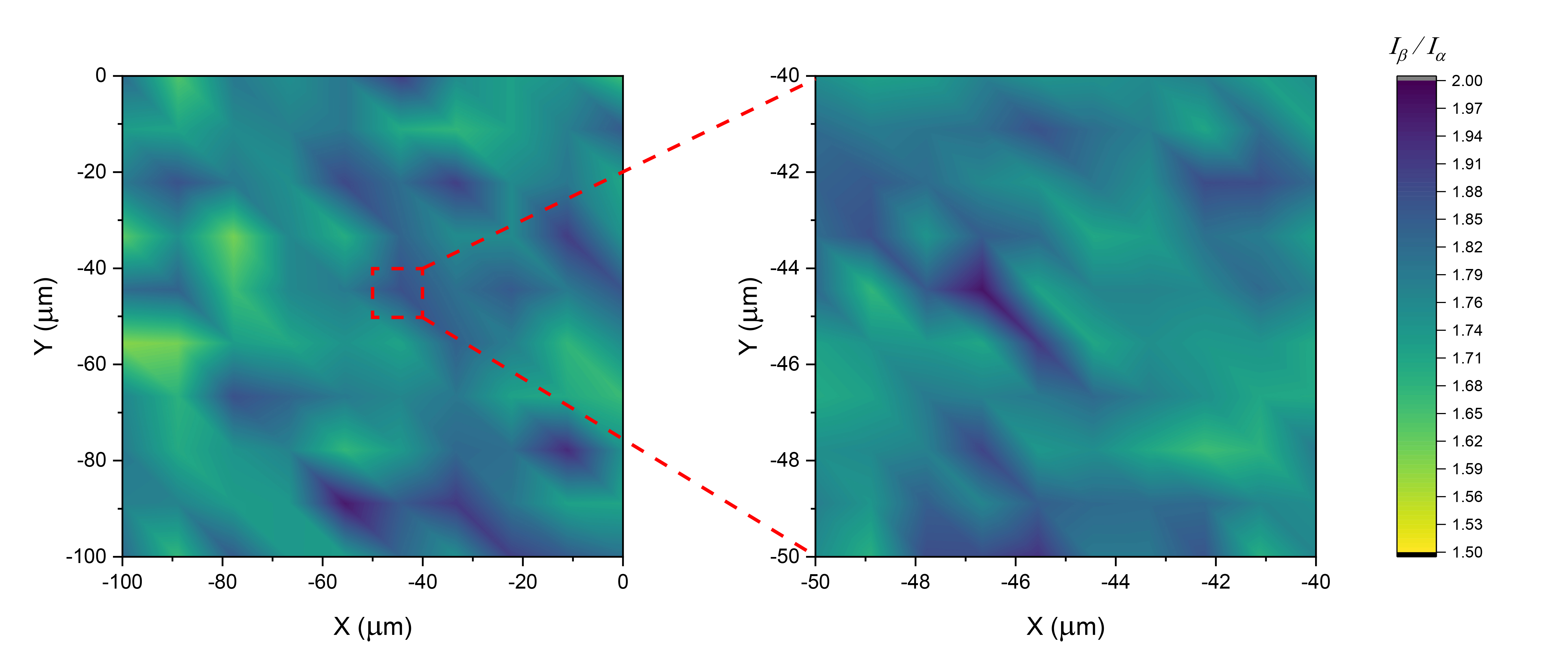}
\caption{Confocal Raman microscopy map of a $100\ \upmu$m $\times 100\ \upmu$m area on the surface of a EM-SRBW PVDF film, showing local distributions of the $\alpha$- and $\beta$-phases, quantified by the relative intensity between the Raman $\beta$-phase peak at 839 cm$^{-1}$ and the $\alpha$-phase peak at 794 cm$^{-1}$ ($I_\beta/I_\alpha$). The magnified inset represents an area of  10 $\upmu$m $\times$  10 $\upmu$m.}
\label{Sfigure2}
\end{figure}
\newpage

\begin{figure}[H]
\centering
\hspace*{-1cm}
\includegraphics[width=0.85\textwidth]{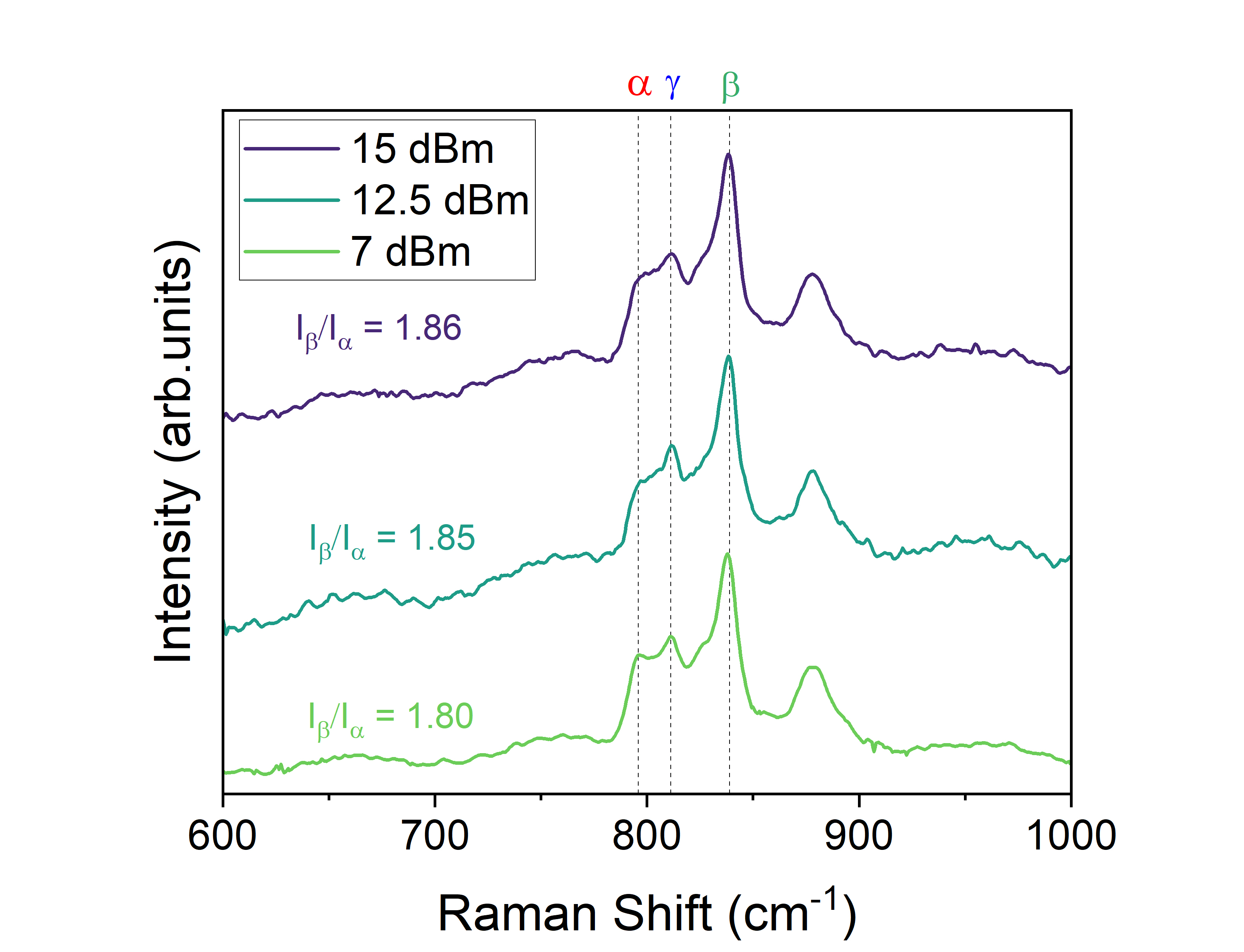}
\caption{Variation in the phase composition of EM-SRBW PVDF films synthesized at different input SRBW powers, as quantified using Raman spectroscopy.}
\label{Sfigure3}
\end{figure}
\newpage

\begin{figure}[H]
\centering
\hspace*{-1cm}
\includegraphics[width=0.85\textwidth]{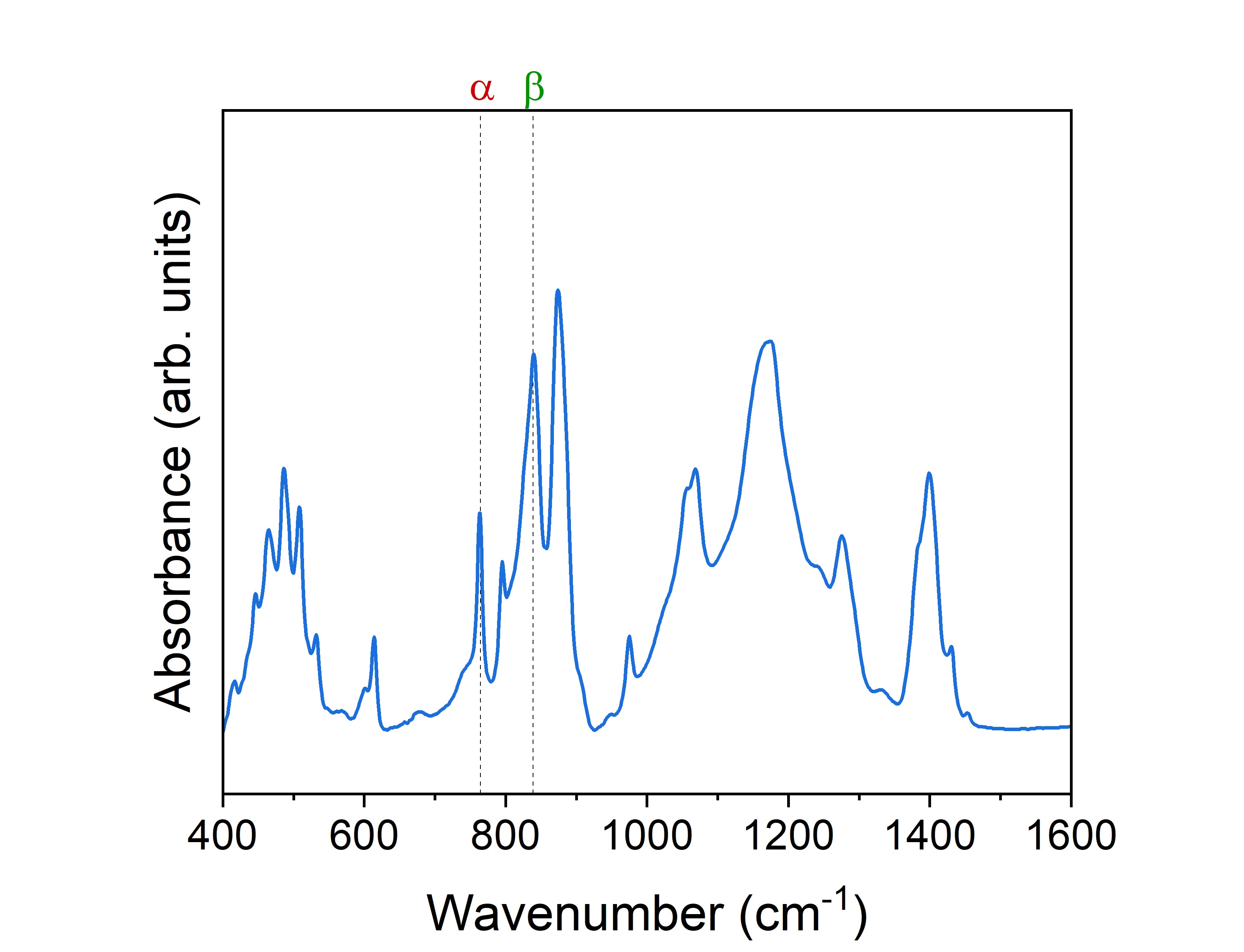}
\caption{Fourier Transform infrared (FTIR) spectra for the 6 $\upmu$m commercially-poled PVDF film. The FTIR peaks for the $\alpha$- and $\beta$-phases are located at 763 and 841 cm$^{-1}$, respectively.}
\label{Sfigure4}
\end{figure}
\newpage

\begin{figure}[H]
\centering
\hspace*{-1cm}
\includegraphics[width=0.8\textwidth]{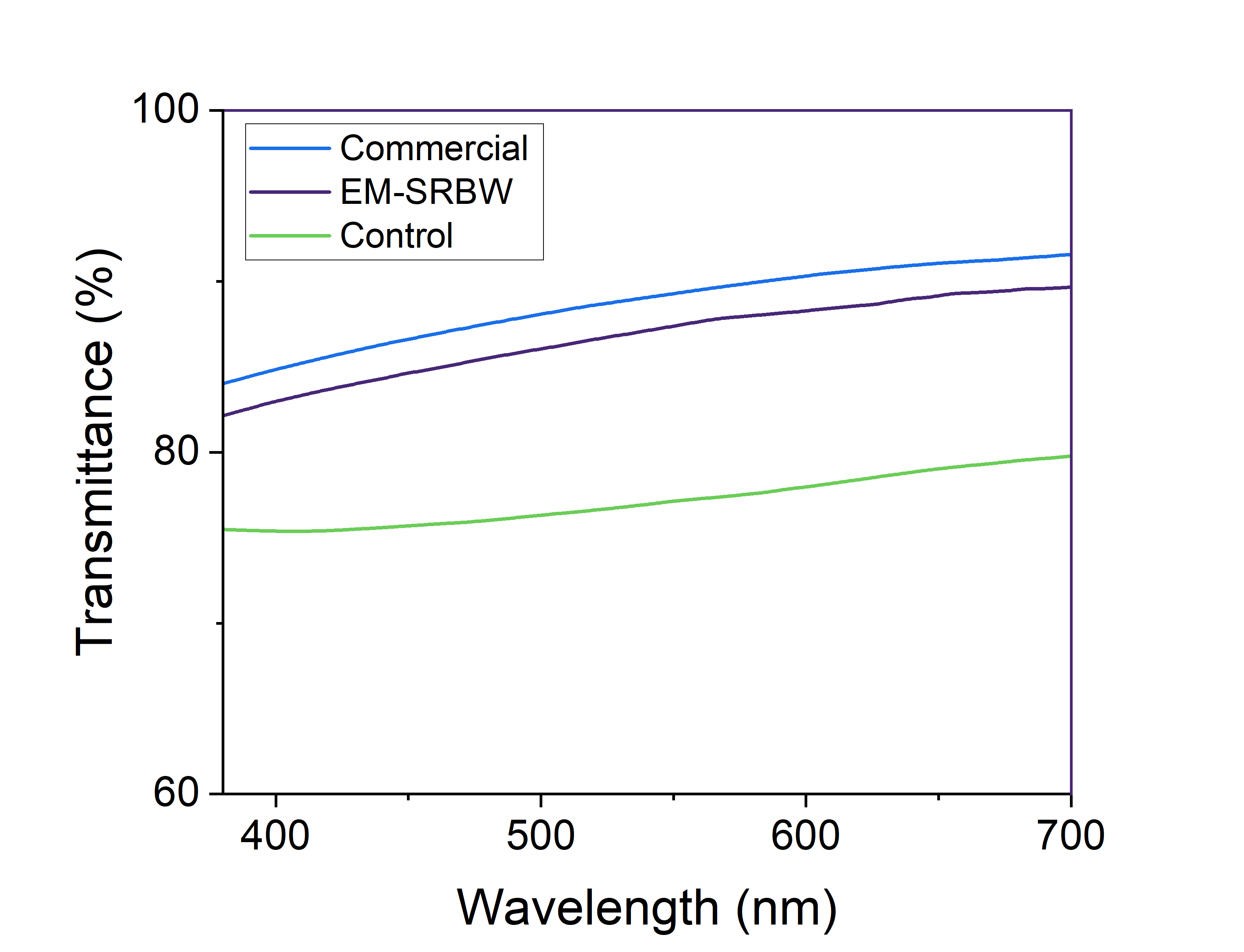}
\caption{Total light transmittance of the control, commercial-poled and EM-SRBW PVDF films (6 $\upmu$m thick).}
\label{Sfigure1}
\end{figure}
\newpage

\begin{figure}[H]
\centering
\hspace*{-1cm}
\includegraphics[width=1.2\textwidth]{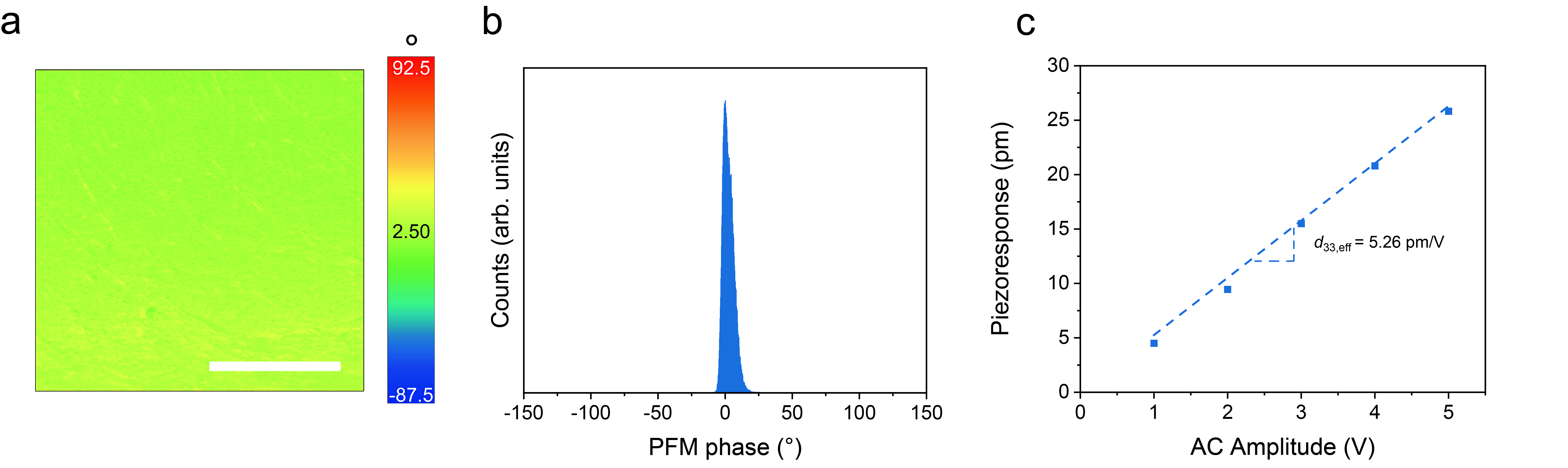}
\caption{Local polarisation of $6\ \upmu$m thick commercially-poled PVDF films. Piezoresponse force microscopy (PFM) (a) phase image, and, (b) phase histogram. (c) Piezoelectric response and corresponding piezoelectric coefficient ($d_{33,{\rm eff}}$).}
\label{Sfigure5}
\end{figure}
\newpage

\begin{figure}[H]
\centering
\hspace*{-1cm}
\includegraphics[width=1.1\textwidth]{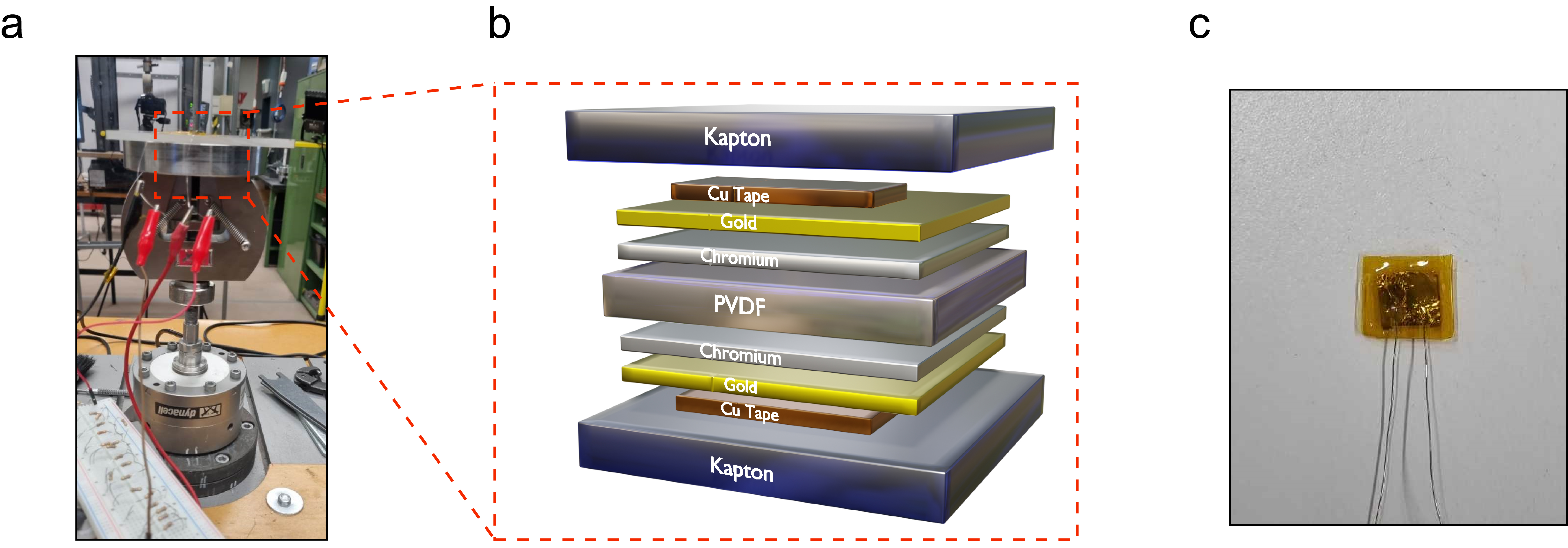}
\caption{(a) Setup for the macroscopic piezoelectric response measurements of the various piezoelectric energy harvesting (PENG) devices using a dynamic testing instrument (Electropuls E3000; Instron, Norwood, MA, USA). (b) Schematic of the component layout for the PENG devices. (c) Image of an example PENG device.}
\label{Sfigure6}
\end{figure}
\newpage

\begin{figure}[H]
\centering
\hspace*{-2cm}
\includegraphics[width=1.1\textwidth]{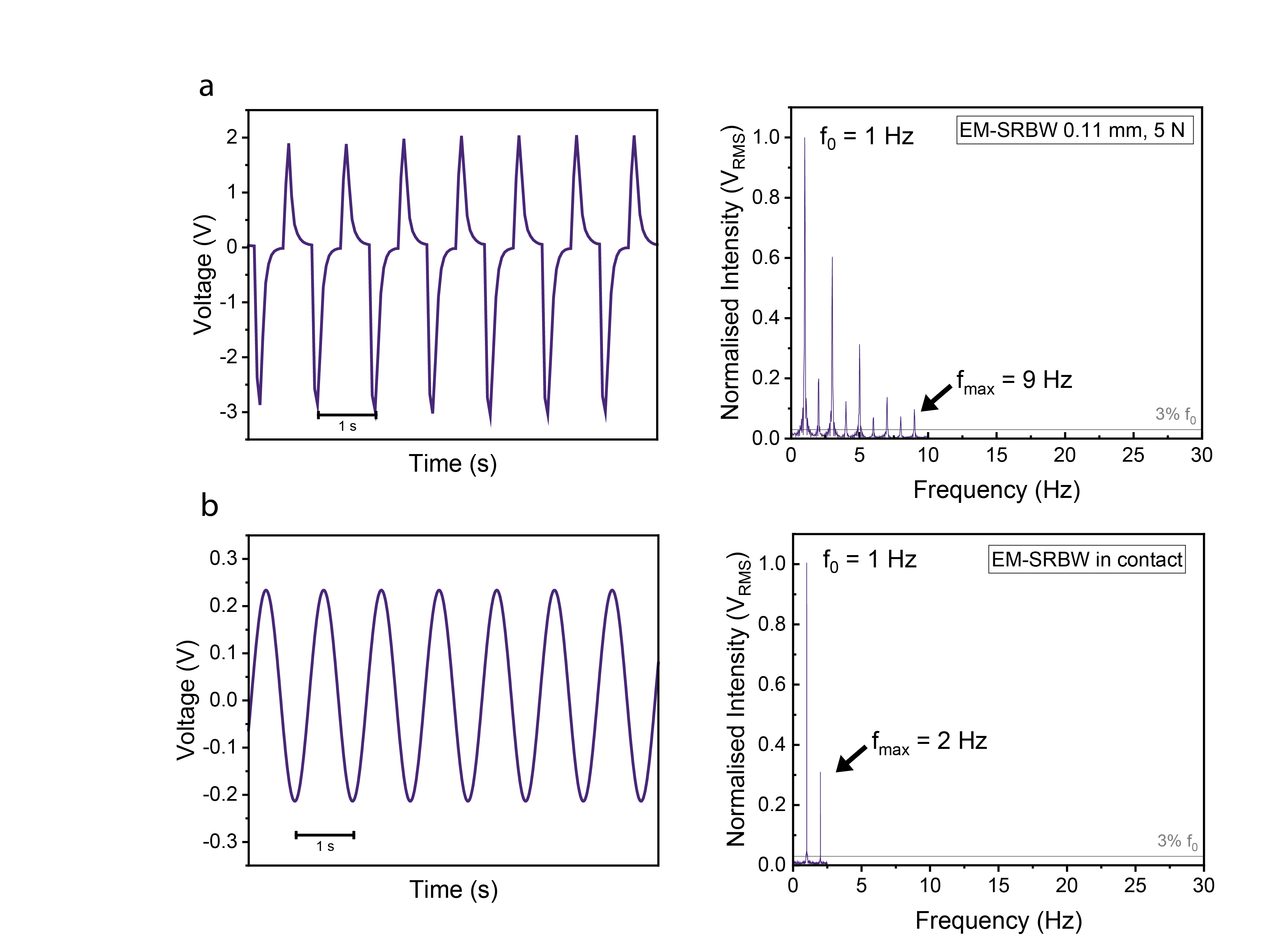}
\caption{(a) Time domain waveforms and corresponding fast Fourier Transform (FFT) spectrum of signals from macroscopic piezoelectric response measurements with 0.11 mm contact separation, and, (b) under in-contact mode, for the EM-SRBW PENG device. }
\label{Sfigure7}
\end{figure}
\newpage

\begin{figure}[H]
\centering
\hspace*{-1cm}
\includegraphics[width=0.85\textwidth]{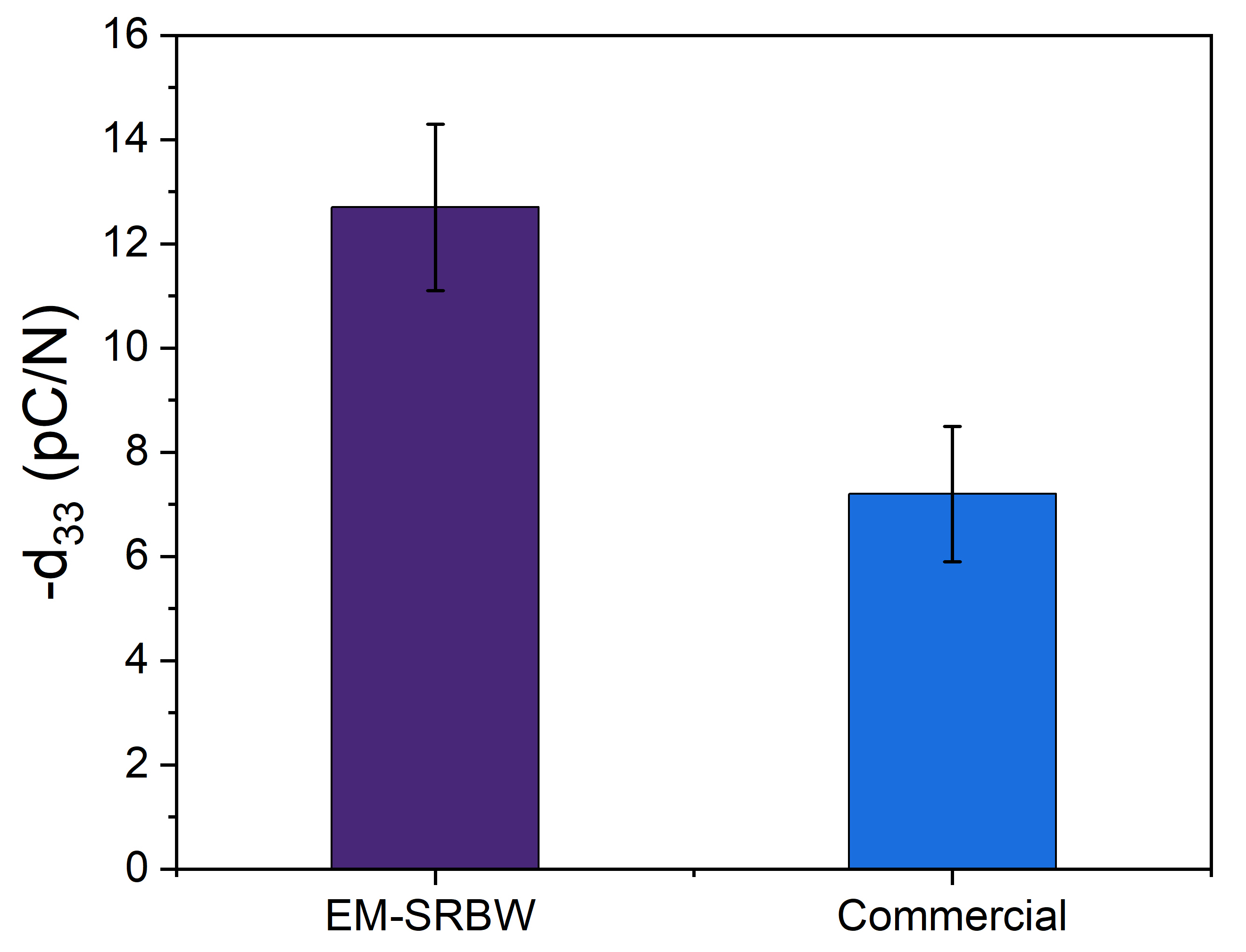}
\caption{Measurement of the piezoelectric charge coefficient ($d_{33}$) for the $6\ \upmu$m thick EM-SRBW and commercially-poled films using a piezoelectric meter with a 2 N static load.}
\label{Sfigure8}
\end{figure}
\newpage

\begin{figure}[H]
\centering
\hspace*{-1cm}
\includegraphics[width=1.1\textwidth]{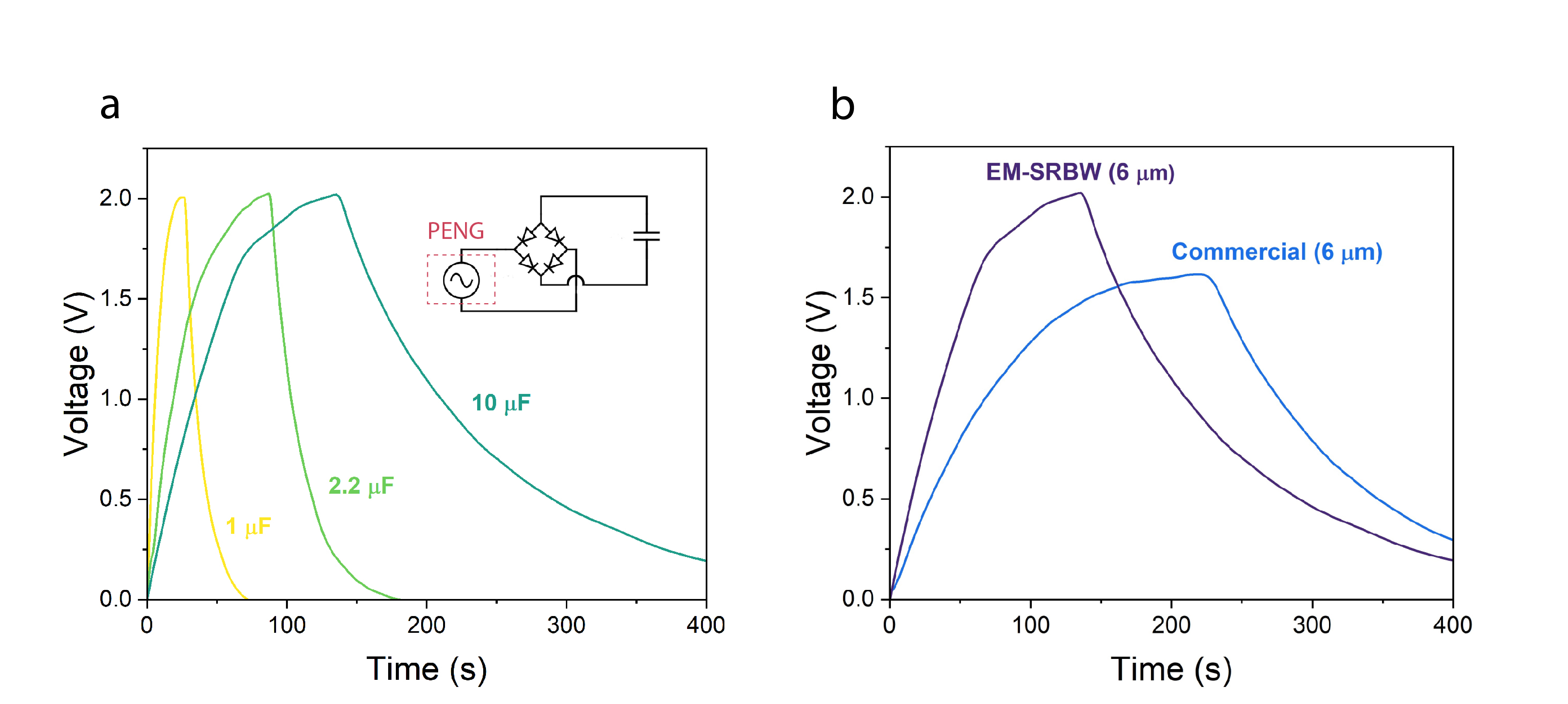}
\caption{(a) Charge--discharge profiles for PENG devices comprising 6 $\upmu$m thick EM-SRBW PVDF films when one capacitor (1, 2.2 and 10 $\upmu$F) was used, and, (b) $6\ \upmu$m thick EM-SRBW and commercially-poled films when a 10 $\upmu$F capacitor was utilised. The inset in (a) shows the variable capacitor being charged from voltages supplied by the PENG through the use of a bridge rectifier.}
\label{Sfigure9}
\end{figure}